\documentclass[10pt, conference, letterpaper, table]{IEEEtran}

\IEEEoverridecommandlockouts
\usepackage{graphicx}

\ifCLASSOPTIONcompsoc
    \usepackage[caption=false, font=normalsize, labelfont=sf, textfont=sf]{subfig}
\else
\usepackage[caption=false, font=footnotesize]{subfig}
\fi
\usepackage{cite}
\usepackage[norelsize,boxruled]{algorithm2e}
\usepackage{url}
\usepackage{mdwmath}     
\usepackage[table]{xcolor}
\usepackage{amsfonts,amsmath,amssymb,amsxtra,amsbsy,floatflt}
\usepackage{indentfirst} 
\usepackage{tabularx}
\usepackage{booktabs}
\usepackage{multirow}
\usepackage{nth}
\usepackage{comment}
\usepackage{epstopdf}
\usepackage{bbold}
\usepackage{hhline}
\usepackage{siunitx}
\usepackage[acronym]{glossaries}
\usepackage{booktabs}
\usepackage[bottom]{footmisc}
\usepackage[bottom]{footmisc}
\usepackage[acronym]{glossaries}
\usepackage{upgreek}
\usepackage{algorithmicx,algpseudocode}
\usepackage{bm}
\usepackage[utf8]{inputenc}
\usepackage{lipsum}
\usepackage{array} 
\usepackage{epsfig}
\usepackage{times}
\usepackage{framed}
\usepackage{enumitem}
\usepackage{float}
\usepackage{etoolbox}
\usepackage{lipsum}
\usepackage{soul}
\usepackage{mathtools}
\usepackage{bbm}
\usepackage{makecell}
\usepackage{pifont}
\usepackage[export]{adjustbox} 
\usepackage{latexsym}
\usepackage{multicol}
\usepackage{eurosym}
\usepackage{xspace}
\usepackage{pdfpages}
\usepackage{verbatim}
\usepackage{stfloats}
\usepackage{blindtext}

{}
{}
{}
{}
{}
{}
{}
\newtheorem{problem}{Problem}

\newcommand{\Compl}{\mbox{$\mathbb{C}$}}
\newcommand{\tran}{\mathrm{T}}
\newcommand{\Exp}{\mathbb{E}}

\newcommand{\rmL}{\mathrm{d}}

\newcommand{\maximize}{\mathrm{maximize}}

\newcommand{\h}{\mathbf{h}}

\newcommand{\s}{\mathbf{s}}

\renewcommand{\v}{\mathbf{v}}
\newcommand{\w}{\mathbf{w}}
\newcommand{\x}{\mathbf{x}}

\newcommand{\G}{\mathbf{G}}
\renewcommand{\H}{\mathbf{H}}

\newcommand{\W}{\mathbf{W}}

\newcommand{\Phib}{\mathbf{\Phi}}
\newcommand{\diag}{\mathrm{diag}}

\newcommand{\herm}{\mathrm{H}}

\renewcommand{\Re}{\mathrm{Re}}

\newacronym{3d}{3D}{three dimensional}
\newacronym[plural=RISs, firstplural=reconfigurable intelligent surfaces (RISs)]{ris}{RIS}{reconfigurable intelligent surface}
\newacronym{ula}{ULA}{uniform linear array}
\newacronym{upa}{UPA}{uniform planar array}
\newacronym{ap}{AP}{access point}
\newacronym{pdf}{pdf}{probability distribution function}
\newacronym{aod}{AoD}{angle of departure}
\newacronym{aoa}{AoA}{angle of arrival}
\newacronym{los}{LoS}{line-of-sight}
\newacronym{pla}{PLA}{planar linear array}
\newacronym{sdp}{SDP}{semidefinite programming}
\newacronym{sdr}{SDR}{semidefinite relaxation}
\newacronym{sre}{SRE}{smart radio environment}
\newacronym{snr}{SNR}{signal-to-noise ratio}
\newacronym{toa}{ToA}{time-of-arrival}
\newacronym{doa}{DoA}{direction-of-arrival}
\newacronym{mmse}{MMSE}{minimum mean squared error}
\newacronym{peb}{PEB}{position error bound}
\newacronym{oeb}{OEB}{orientation error bound}
\newacronym{rss}{RSS}{received signal strength}
\newacronym{ml}{ML}{machine learning}
\newacronym{rmse}{RMSE}{root-mean-square error}
\newacronym{ul}{UL}{uplink}
\newacronym{em}{EM}{electromagnetic}
\newacronym{soa}{SOA}{state-of-the-art}
\newacronym[plural=BSs, firstplural=base stations (BSs)]{bs}{BS}{base station}
\newacronym[plural=UEs, firstplural=user equipments (UEs)]{ue}{UE}{user equipment}

\title{RIShield: Enabling Electromagnetic Blackout in Radiation-Sensitive Environments\\
}
\author{G. Encinas-Lago, M. Rossanese, V. Sciancalepore, M Di Renzo, X. Costa-Perez}

\author{\IEEEauthorblockN{
G. Encinas-Lago\IEEEauthorrefmark{1},
M. Rossanese\IEEEauthorrefmark{1}, 
V. Sciancalepore\IEEEauthorrefmark{1},
Marco Di Renzo\IEEEauthorrefmark{2},
Xavier Costa-Pérez\IEEEauthorrefmark{1}\IEEEauthorrefmark{3}
}                                     % ...
\IEEEauthorblockA{\IEEEauthorrefmark{1}% 1st affiliations
NEC Laboratories Europe GmbH, Heidelberg, Germany}
\IEEEauthorblockA{\IEEEauthorrefmark{2}% 4th affiliations
Université Paris-Saclay, CNRS, Centrale Supélec, Gif-sur-Yvette, France}  
\IEEEauthorblockA{\IEEEauthorrefmark{3}% 5th affiliations
i2CAT Foundation and ICREA, Barcelona, Spain}
\thanks{The research leading to these results has received funding from the EU FP for Research and Innovation Horizon 2020 under Grant Agreements No. 101017011 (RISE-6G), 861222 (MINTS) and 956256 (METAWIRELESS).}}

\renewcommand{\baselinestretch}{0.97}
\begin{document}

\maketitle

\begin{abstract}
Reconfigurable Intelligent Surfaces (RIS) have emerged as a disruptive technology with the potential to revolutionize wireless communication systems. In this paper, we present RIShield, a novel application of RIS technology specifically designed for radiation-sensitive environments. The aim of RIShield is to enable electromagnetic blackouts, preventing radiation leakage from target areas.
We propose a comprehensive framework for RIShield deployment, considering the unique challenges and requirements of radiation-sensitive environments. By strategically positioning RIS panels, we create an intelligent shielding mechanism that selectively absorbs and reflects electromagnetic waves, effectively blocking radiation transmission.

To achieve optimal performance, we model the corresponding channel and design a dynamic control that adjusts the RIS configuration based on real-time radiation monitoring. By leveraging the principles of reconfiguration and intelligent control, RIShield ensures adaptive and efficient protection while minimizing signal degradation.
Through full-wave and ray-tracing simulations, we demonstrate the effectiveness of RIShield in achieving significant electromagnetic attenuation. Our results highlight the potential of RIS technology to address critical concerns in radiation-sensitive environments, paving the way for safer and more secure operations in industries such as healthcare, nuclear facilities, and defense.
\end{abstract}

\glsresetall

\section{Introduction}
\label{sec:introduction}

With the increasing need for privacy and security in radiation-sensitive environments, such as medical facilities, research laboratories, and governmental institutions, there is a growing demand for innovative technologies that can effectively isolate specific areas and prevent electromagnetic leakage. Literature approaches suggest installing metal barriers provided with electromagnetic absorbers to limit the incoming signal and guarantee isolation. However, such solutions may prevent flexibility and versatility in highly dynamic environments, where shields are required for a short time according to on-demand events.

In this context, Reconfigurable Intelligent Surfaces (RIS) have emerged as a promising solution, offering unprecedented opportunities for shielding and protecting sensitive spaces in a dynamic fashion~\cite{Awarkeh22}. Such emerging technology exhibits its peculiar property of altering programmatically the surrounding propagation environment: impinging signals can be refracted, absorbed, or reflected upon a specific RIS configuration. This makes RIS the perfect candidate to gain control of the propagation environment --- always conceived as a black-box --- thereby powering a new wireless revolution. Ultimately, RISs are designed to be low-complex, nearly passive, and inexpensive devices~\cite{rossanese2022designing}. This would extremely increase the technology penetration rate, envisioning a massive diffusion of such novel devices in the next few years.  

\begin{figure}[t!]
    \centering
    \includegraphics[width=\columnwidth]{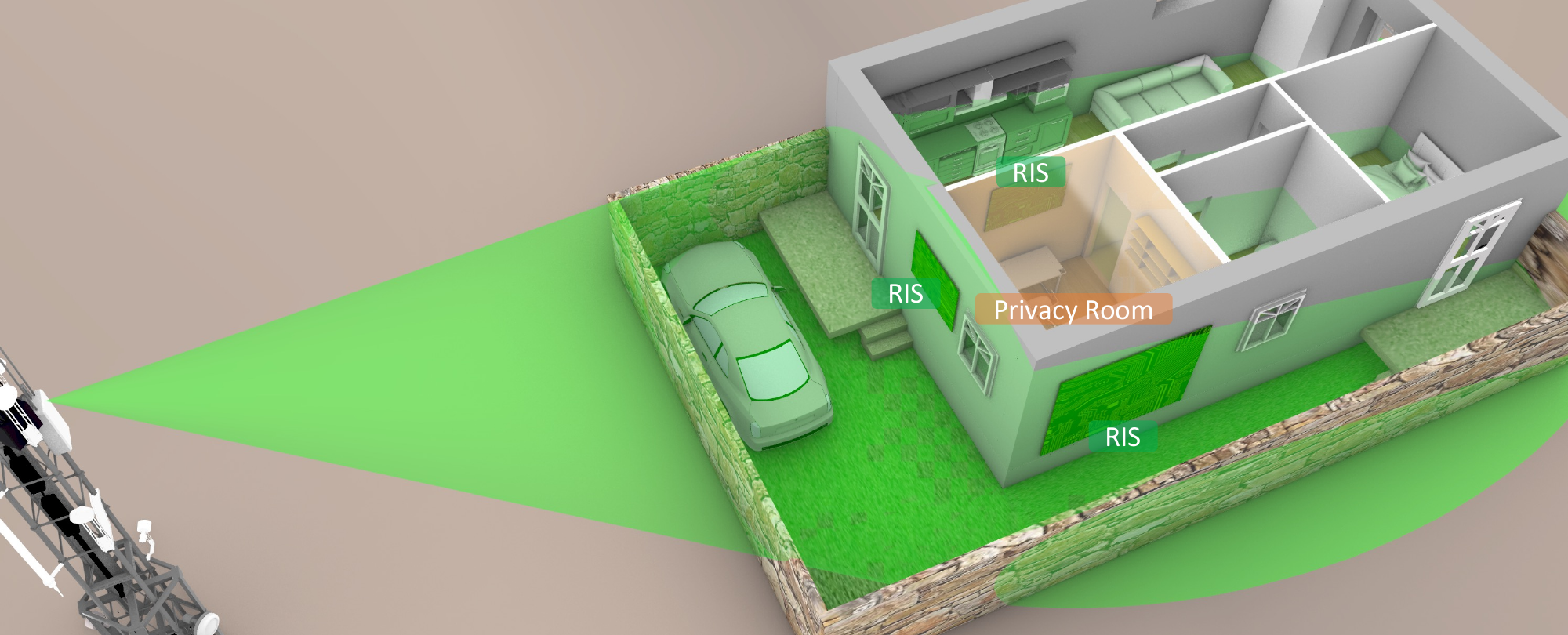}
    \caption{RIS used to guarantee privacy and isolation in real environments.}
    \label{fig:scenario}
\end{figure}

Only a few state-of-the-art proposals have looked at the problem. In particular, ~\cite {wang2023} mitigates cross-system interference for secure wireless applications. The authors introduced absorptive RIS (ARIS) to modify both the phase and modulus of the impinging signal by absorbing a portion of the signal energy. Also, \cite{Huang2020} introduced the concept of Holographic Multiple Input Multiple Output Surface (HMIMOS) where the RIS is applied to specific scenarios where EM-Field Absorption is required. However, none of the available literature works has proposed a practical solution to exploit the RIS to isolate specific areas/rooms in both outdoor and indoor scenarios.

Motivated by such an innovative and unprecedented characteristic, in this paper we design and present RIShield, a pioneering application of RIS technology specifically designed to enable electromagnetic blackouts and ensure privacy in radiation-sensitive environments. In particular, RIShield leverages the unique capabilities of RIS to selectively absorb and reflect electromagnetic waves, effectively creating a dynamic and intelligent shield that isolates and safeguards specific rooms or areas from unwanted radiation leakage on-demand, as shown in Fig.~\ref{fig:scenario}.

By strategically deploying RIS panels within the environment, RIShield empowers administrators and operators with the ability to dynamically control and manage electromagnetic propagation. This level of control not only guarantees privacy by preventing sensitive information from being intercepted but also minimizes the risk of interference with external electromagnetic signals.
To achieve optimal performance and adaptability, we propose the development of an advanced optimization solution that dynamically configures the RIS system based on real-time monitoring of radiation levels. These algorithms ensure that RIShield provides effective shielding while minimizing signal degradation and accommodating changing requirements within the radiation-sensitive environment.
Through a combination of simulation studies and experimental evaluations, we aim to demonstrate the efficacy of RIShield in achieving electromagnetic blackout and guaranteeing privacy in radiation-sensitive environments. The preliminary results of this analysis will provide valuable insight into the potential of RIS technology as a versatile and efficient solution for addressing the privacy and security challenges faced in radiation-sensitive settings.
Overall, RIShield opens up new horizons in shielding technology, enabling the establishment of radiation-secure spaces and laying the foundation for enhanced privacy and confidentiality in critical sectors where electromagnetic isolation is of utmost importance.

\subsection{Notation} 
We denote vectors and matrices using bold-face lower-case and upper-case letters, respectively. We use italic letters to denote scalars. We use $\Compl^n$ and $\Compl^{n\times m}$ to represent the sets of $n$-dimensional complex vectors and $m\times n$ complex matrices, respectively. Vectors are denoted as column vectors. Subscripts represent an element in a vector and superscripts elements in a sequence. For example, $\x^{(t)} = [x^{(t)}_1, \dots, x^{(t)}_n]^{\tran}$ is a vector from $\Compl^n$ and $x^{(t)}_i$ is its $i$th component. In addition, operation $(\cdot)^{\tran}$ represents the transpose operator, $\otimes$ stands for Kronecker product while $(\cdot)^{\herm}$ denotes the Hermitian transpose operation. Finally, $\|\cdot\|$ and $\|\cdot\|_{\mathrm{F}}$ denote the L2-norm of a vector and Frobenius norm of a matrix, respectively.

\section{System Model}
\label{sec:system_model}

For the sake of simplicity, we model Reconfigurable Intelligent Surfaces (RISs) as reflectarrays composed of multiple antenna elements placed following a grid configuration where the inter-distance among each pair of elements is equal to $\lambda/2$~\cite{SARIS_WCL_2023}. This prevents us from experiencing the grating lobe effect or mutual coupling among the elements~\cite{renzo20}.
Each antenna element can be only configured to reflect or absorb the impinging signal in a passive way: no active components are placed on the RIS board.

Let us consider a conventional scenario where a transmitter is equipped with $\Sigma$ antennas and the set $\mathcal{U}=\{1,\ldots U\}$ denotes the receivers provided with a single antenna. The connection is established using a set of Reflecting Intelligent Surfaces (RISs) installed on the buildings or internal walls. Each RIS is equipped with $N$ equivalent antenna elements. 

For downlink data transmission, the transmitter (tx) communicates with each receiver $u$ through a direct link represented by $\h_{\rmL,k} \in \Compl^{\Sigma\times 1}$. This link consists of a line-of-sight (LoS) path with a length of $d_u$ and an angle of departure (AoD) $\theta_u$, when the latter exists, along with a multipath non-line-of-sight (NLoS) component. Furthermore, the transmitter can utilize a combined link to the RIS, denoted by $\G \in \Compl^{N\times\Sigma}$. 

The RIS reflects the incoming signal towards the receiver through the channel $\h_u \in \Compl^{N\times 1}$. The $\h_u$ channel can be decomposed into the LoS tx-RIS path with length $d_{1,u}$, having an AoD from the transmitter and an angle of arrival (AoA) at the RIS denoted by $\psi_{D}$ and $\psi_{A}$, respectively. This is accompanied by a set of scattered NLoS paths and the RIS-receiver~$u$ link, which comprises a LoS path with length $d_{2,u}$ and AoD $\psi_u$, if available, and a multipath NLoS component.
Lastly, signals that are reflected more than 1 time by the RIS are neglected to account for high path loss~\cite{Wu18} as considered in the literature~\cite{Zha19_2, Mis19}.

All channels in the system are modeled as quasi-static flat-fading, which means they remain constant during the transmission time of a codeword. Additionally, we assume that the transmitter has access to perfect Channel State Information (CSI), i.e., it knows the channel vectors ${\h_{\rmL,u}}{u=1}^U$, $\G$, and ${\h{u}}_{u=1}^U$. The transmitter operates using the time division duplex mode, allowing the uplink and downlink channels to be reciprocal. As a result, the downlink physical channel can be estimated using the uplink training from any single receiver via a separate control channel\,\footnote{Imperfect channel information requires a channel estimation process, which has been already explored in~\cite{He19, Mis19} and thus is out of the scope of this analysis.}.

Each receiver $u$ receives the sum of two contributions, such as a direct path from the transmitter and a reflected path that has been steered by the RIS. Hence, we can write the received signal at receiver $u$ as the following
\begin{align}
y_u & = \left(\h_u^{\herm}\Phib\G + \h_{\rmL,u}^{\herm}\right)\W\s + n_u \in \Compl \\
& = \left(\h_u^{\herm}\Phib\G + \h_{\rmL,u}^{\herm}\right)\w_u s_u \nonumber \\
& \phantom{= } + \sum_{j\neq u} \left(\h_u^{\herm}\Phib\G + \h_{\rmL,u}^{\herm}\right)\w_j s_j + n_u,
\end{align}
where  
\begin{align}
    \Phib & = \begin{bmatrix}
    \alpha_1e^{j\phi_1} & 0 & \ldots & 0 \\
    \vdots & \alpha_2e^{j\phi_2} & \ldots & \vdots \\
    0 & \ldots & \ddots &  \alpha_Ne^{j\phi_N}
    \end{bmatrix}
\end{align}
with $\phi_i \in [0,2\pi)$ and $|\alpha_i|^2\leq 1, \,\, \forall i$ represents the phase shifts and amplitude attenuation introduced by the RIS, $\W = [\w_1,\ldots,\w_U] \in \Compl^{\Sigma\times U}$ is the transmit precoder at the transmitter, $\s = [s_1,\ldots,s_U]^{\tran}\in \Compl^{U\times 1}$ is the transmit symbol vector with $\Exp[|s_u|^2] = 1, \, \forall u$, and $n_u$ is the noise term distributed as $\mathcal{CN}(0,\sigma_n^2)$. 

Now, let us assume a single receiver decoding the signal, the system sum rate can be defined as follows
\begin{align}\label{eq:srate}
    R\! \triangleq \! \sum_u \log_2\bigg(\! 1\!+\! \frac{|(\h_u^{\herm}\Phib\G + \h_{\rmL,u}^{\herm})\w_u|^2}{\sum_{j\neq u} |(\h_u^{\herm}\Phib\G + \h_{\rmL,u}^{\herm})\w_j|^2 + \sigma_n^2}\bigg).
\end{align}

In our reference scenario, the RIS is used to limit the incoming signal for specific areas or receivers. Therefore, we formulate the problem aiming at optimizing the overall system performance of the considered RIS-empowered network in terms of system sum rate as per Eq.~\eqref{eq:srate}. This would allow us to directly optimize the precoding strategy at the transmitter and the reflection coefficient at the RIS. However, given the complexity of such an expression, we aim at optimizing the sum-mean-square-error (SMSE) over all receivers, known to be related to Eq.~\eqref{eq:srate}, as reported in~\cite{Mursia2020}. %In particular, the relation between minimum mean squared error (MSE) of receiver $u$ and maximum SINR of receiver $u$ holds for linear filters. 

Hence, we reformulate our problem by looking at SMSE as a means to reduce signal leakage and improve privacy and security in specific locations, while maintaining the connectivity in the other spaces. Considering all the receivers, $u$, placed within the room or area where we would like to guarantee privacy and EMF isolation, we aim at maximizing the overall SMSE for all those receivers. 
The received MSE of receiver $u$ in a specific location is given by
\begin{align}
\mathrm{MSE}_u & = \Exp[|y_u - s_u|^2]\\
& = |(\h_u^{\herm}\Phib\G + \h_{\rmL,u}^{\herm})\w_u - 1|^2 + \nonumber \\
& \phantom{= }\sum_{j\neq u} |(\h_u^{\herm}\Phib\G + \h_{\rmL,u}^{\herm})\w_j|^2 + \sigma_n^2 \\
& = \sum_j |(\h_u^{\herm}\Phib\G + \h_{\rmL,u}^{\herm})\w_j|^2 \nonumber \\
& \phantom{= }- 2 \, \Re\{(\h_u^{\herm}\Phib\G + \h_{\rmL,u}^{\herm})\w_u\} + 1 + \sigma_n^2.
\end{align}
whereas the receive SMSE over all receivers ($\forall u$) is expressed as the following
\begin{align}
\mathrm{SMSE} & = \sum_u \mathrm{MSE}_u \\
& = \sum_u \sum_j |(\h_u^{\herm}\Phib\G + \h_{\rmL,u}^{\herm})\w_j|^2 \nonumber \\
& \phantom{= }- 2 \sum_u \Re\{(\h_u^{\herm}\Phib\G + \h_{\rmL,u}^{\herm})\w_u\}\!+\! U(1 + \sigma_n^2). \label{eq:SMSE}
\end{align}

For ease of presentation, let us define
\begin{align}\label{eq:v}
 \v = [\alpha_1e^{-j\phi_1},\ldots,\alpha_Ne^{-j\phi_N},1]^{\tran}\in \Compl^{N+1\times 1},   
\end{align}
and 
\begin{align}\label{eq:Hb}
\bar{\H}_u \triangleq \begin{bmatrix}
\mathrm{\diag}(\h_u^{\herm}) \G \\
\h_{\rmL,u}^{\herm}
\end{bmatrix}\in \Compl^{N+1\times \Sigma},
\end{align}
such that $\Phib = \mathrm{diag}(\v[1:N]^{\herm})$ and
\begin{align}
    (\h_{u}^{\herm}\Phib\G + \h_{\rmL,u}^{\herm})\w_j = \v^{\herm} \bar{\H}_u \w_j \quad \forall u,j.
\end{align}
Therefore, we can formulate our optimization problem as the following
\begin{problem}[SHIELD]
\begin{align}
    & \displaystyle \underset{\v ,\W}{\maximize} && \displaystyle \sum_u \sum_j |\v^{\herm}\bar{\H}_u\w_j|^2 - 2 \sum_u \Re\{\v^{\herm}\bar{\H}_u\w_u\} & \nonumber\\
    & \mathrm{subject~to} && |v_i|^2 \leq 1, \quad i=1,\ldots,N; \nonumber \\
    &  && v_{N+1} = 1; \nonumber\\
    &  && \|\W\|^2_{\mathrm{F}} \leq P; \nonumber
\end{align}
\end{problem}
\begin{figure}[t]
\begin{minipage}[t]{0.495\linewidth}
\includegraphics[width=\linewidth]{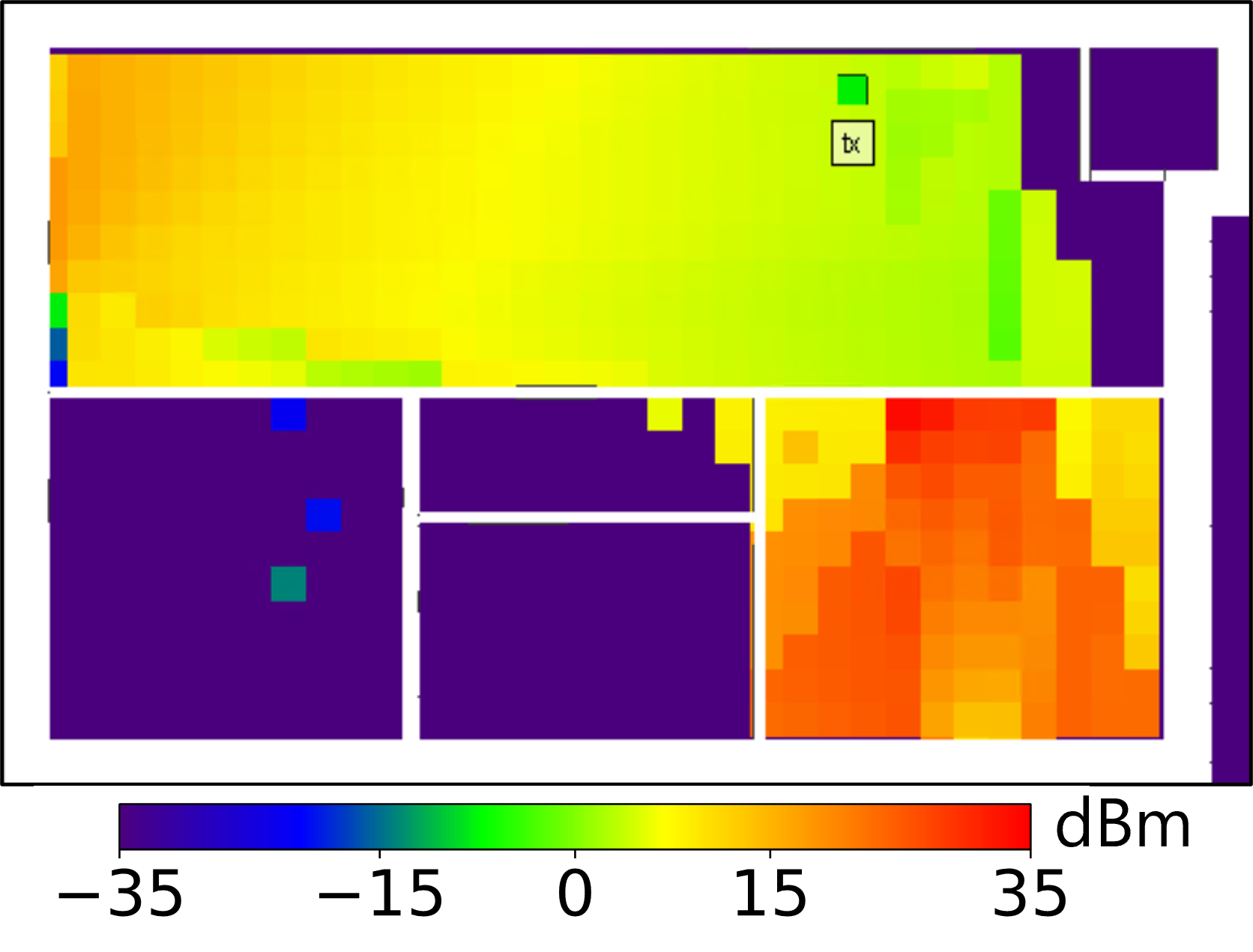}
\caption{Conventional 2D indoor scenario (4 rooms) with 1 indoor transmitter. The signal propagates mostly in the entire considered area through walls.}
\label{fig:baseline2D_bar}
\end{minipage}
\hfill
\begin{minipage}[t]{0.495\linewidth}
\includegraphics[width=\linewidth]{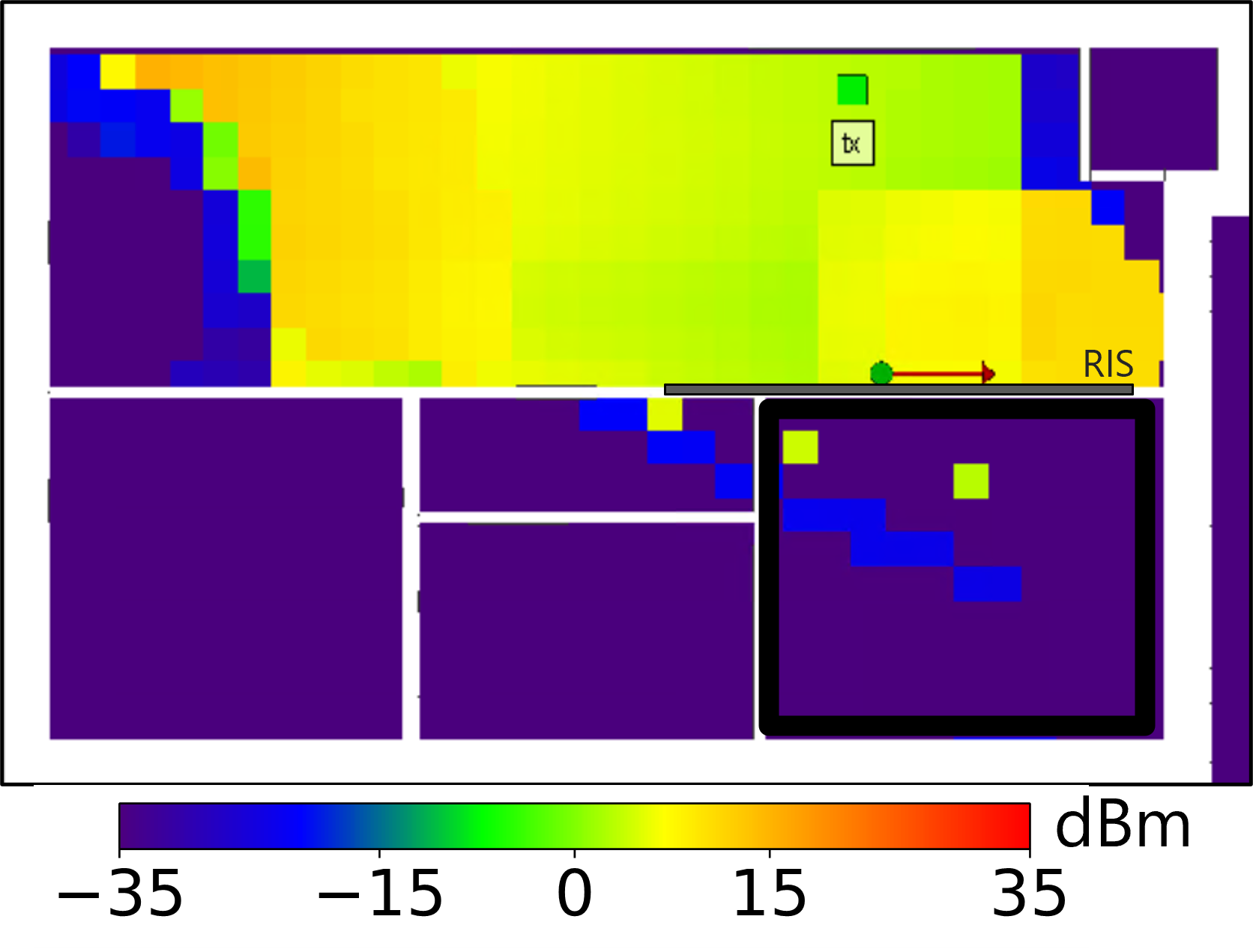}
\caption{Conventional 2D indoor scenario (4 rooms) with 1 indoor transmitter and RIS enabled as a shield. The right-hand side room is isolated.}
\label{fig:RIS2D_bar}
\end{minipage}
\vspace{-4mm}
\end{figure}

\begin{figure}[t]
\begin{minipage}[t]{0.495\linewidth}
\includegraphics[width=\linewidth]{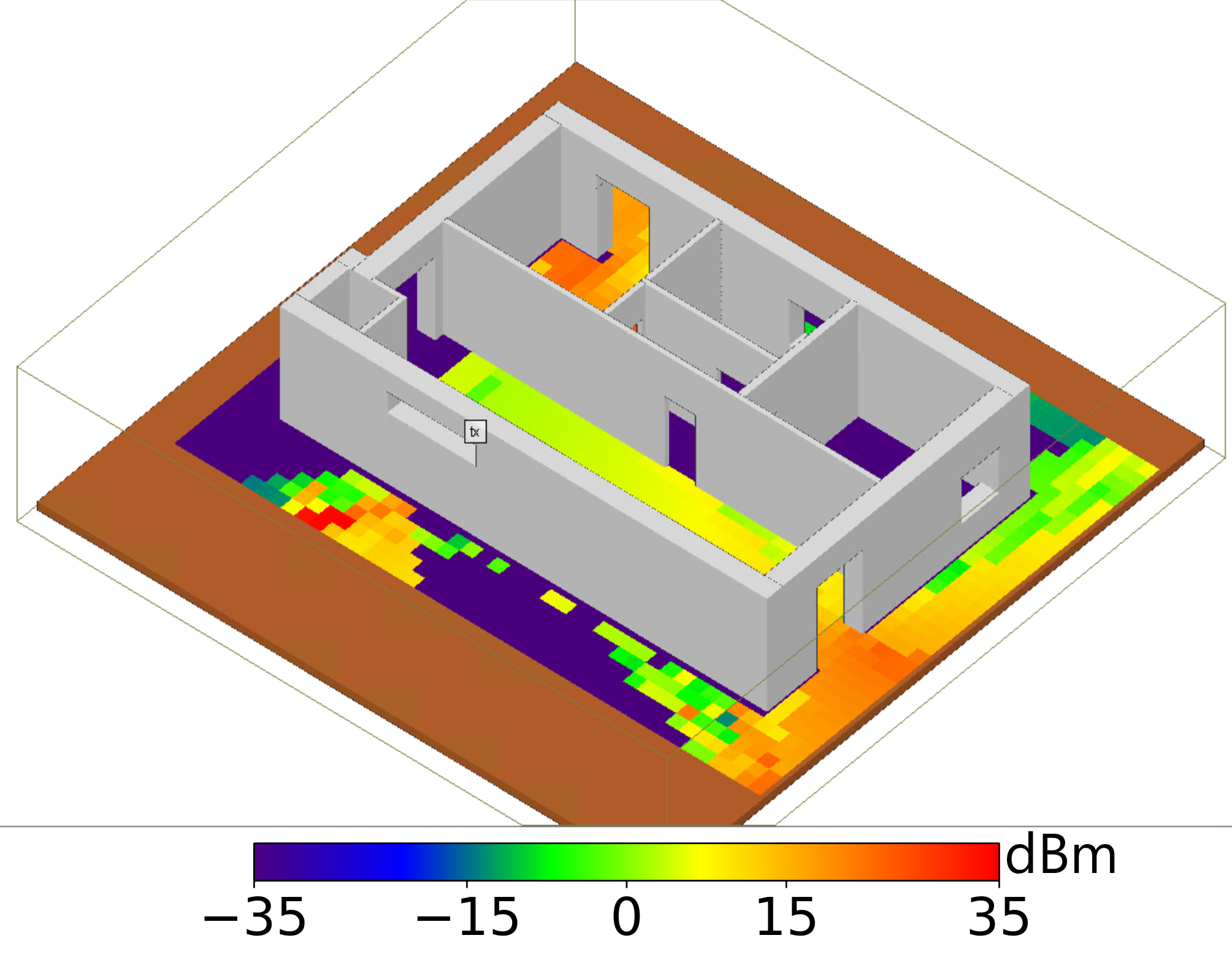}
\caption{Conventional 3D indoor scenario (4 rooms) with 1 indoor transmitter. The signal propagates mostly in the entire considered area through walls.}
\label{fig:baseline3D_bar}
\end{minipage}
\hfill
\begin{minipage}[t]{0.495\linewidth}
\includegraphics[width=\linewidth]{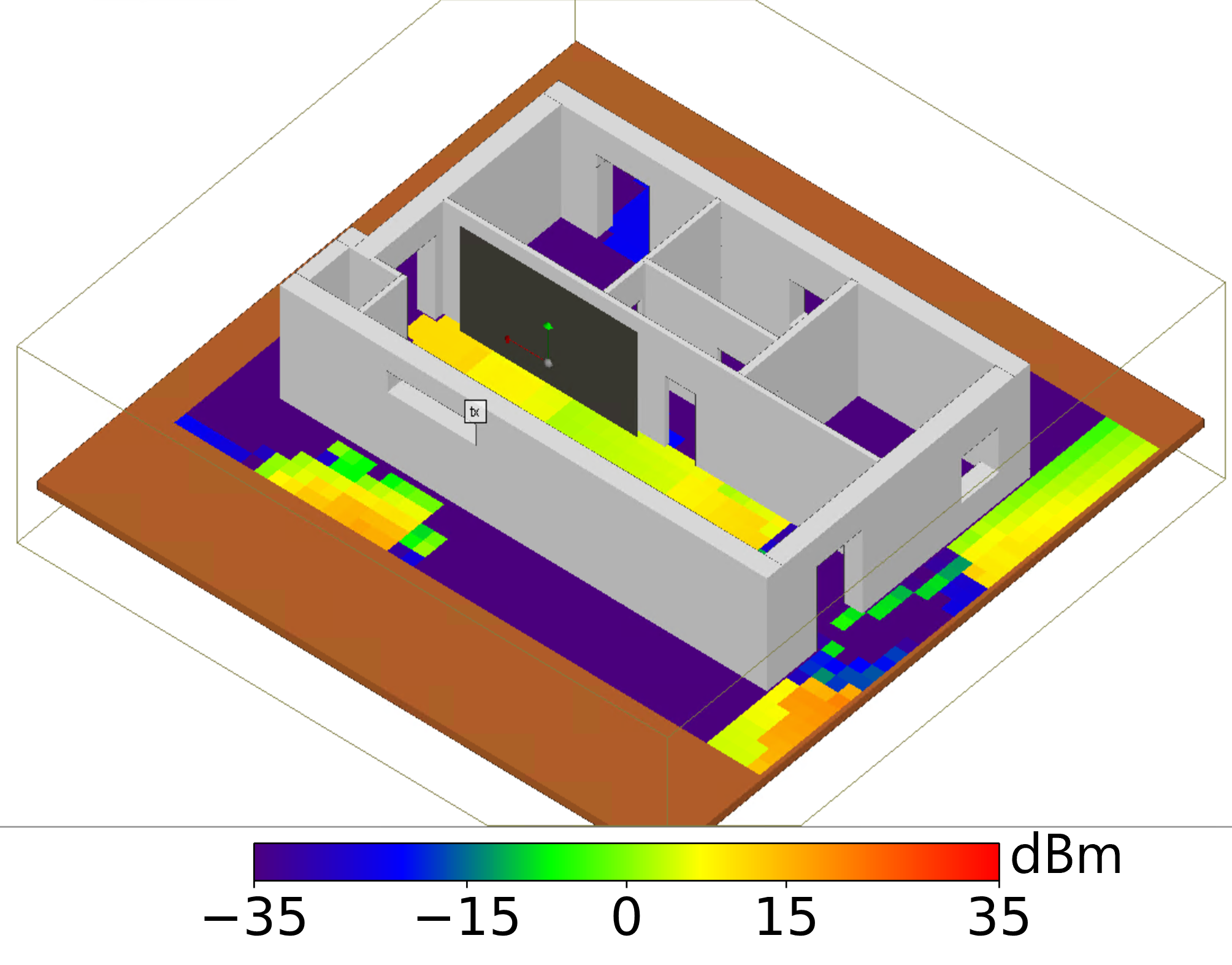}
\caption{Conventional 3D indoor scenario (4 rooms) with 1 indoor transmitter and RIS enabled as a shield. The right-hand side room is isolated.}
\label{fig:RIS3D_bar}
\end{minipage}
\end{figure}
\noindent with $\v$ defined in Eq.~\eqref{eq:v} and $P$ the available transmit power at the transmitter. We have introduced the constraint $|v_i|^2\leq 1$ to ensure the passive assumption, i.e., the $i$-th RIS element does not amplify the incoming signal. While this proposed formulation is not convex, the semidefinite relaxation is used to solve the problem.
% While this proposed formulation is convex in the two optimization variables $\v$ and $\W$, separately, we preliminary fix the $\W$, i.e., the transmit precoder, assuming it as input of our problem. This results in an easier problem where only the RIS configuration ($\v$) is treated as an optimization variable.

\section{Numerical Results}
\label{sec:results}
When dealing with the RIS configuration $\v$ as described in \emph{Problem 1~(SHIELD)}, our solution will trigger a specific phase to be applied to each antenna element. This will turn each RIS element into a reflective or absorbing cell. 
% Given the convexity property, we solve the problem of obtaining the optimal configuration that minimizes the SMSE over all receivers within a specific area. 
To provide a complete overview of our RIShield solution, we test our framework using Remcom WirelessInSite~\cite{WirelessInsite}, a commercial software that includes a set of valid RF propagation models thereby providing 3D ray-tracing, fast ray-based methods, and empirical models for the analysis of site-specific radio wave propagation and wireless communication systems. 

\subsection{WirelessInSite Simulations}
We test a real-environment topology considering an indoor scenario where the transmitter is assumed to be an access point operating at $2.4$GHz with a transmit power $P=20$dbm. The scenario shows a $60$ square meters apartment including $4$ different rooms with $2.7$m height, external walls of $41$cm thick, and internal walls of $10-15$cm thick. The transmitter is placed in the bigger room as shown in Fig.~\ref{fig:baseline2D_bar}.

When no RIS is installed in the apartment, the signal propagates through the entire scenario while being (slightly) attenuated by the internal walls. It can be noted that rooms have doors that allow the signal to easily get into the room. In particular, the right-hand side room in Fig.~\ref{fig:baseline2D_bar} and its $3$-dimensional correspondence in Fig.~\ref{fig:baseline3D_bar}, is strongly irradiated when the door is open.
However, when a fully-absorbing RIS is placed on the wall, the signal propagation is strongly attenuated. In Fig.~\ref{fig:RIS2D_bar} the right-hand side room gets almost-zero signal due to the RIS placed on the wall (as shown in Fig.~\ref{fig:RIS3D_bar}). 

Please note that the room is not completely isolated, as the RIS is only installed on a specific side of the room. The transmitter signal is still able to barely reach the (users inside the) room through the door on the right. Nonetheless, most of the EM Field exposure has been denied with a simple fully-absorbing RIS on the wall.

While this behavior is straightforward and expected, it unveils a two-fold message: $i$) RIS size, orientation and placement have a direct impact on the overall isolation guarantees and, $ii$) a specific RIS configuration (i.e., $\v$ in \emph{Problem 1~(SHIELD)}) can influence the signal spreading and, in turn, the privacy and security application.

\begin{figure*}[t]
\begin{minipage}[t]{0.325\linewidth}
\centering
\includegraphics[width=0.6\linewidth]{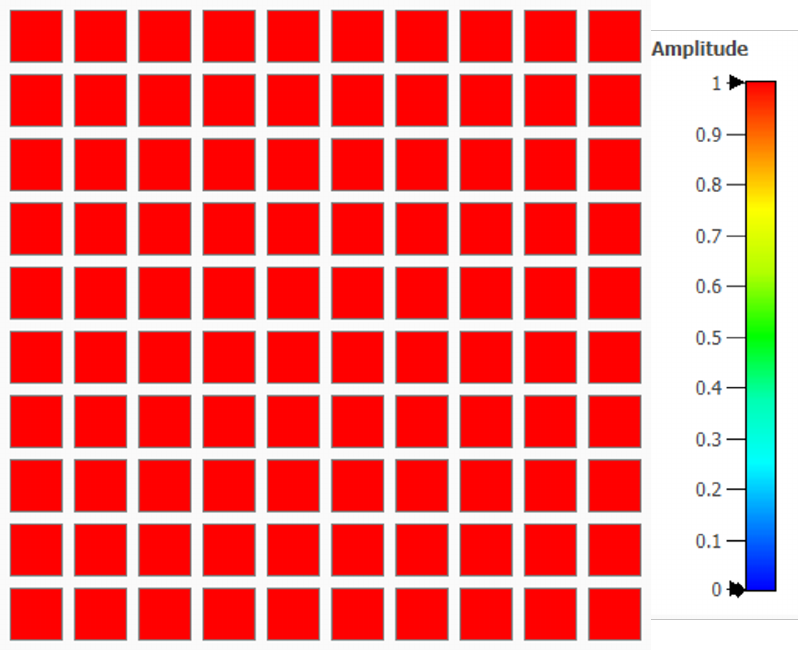}
\caption{RIS configuration to enable full reflection.}
\label{fig:fullRefl}
\end{minipage}
\hfill
\begin{minipage}[t]{0.325\linewidth}
\centering
\includegraphics[width=0.6\linewidth]{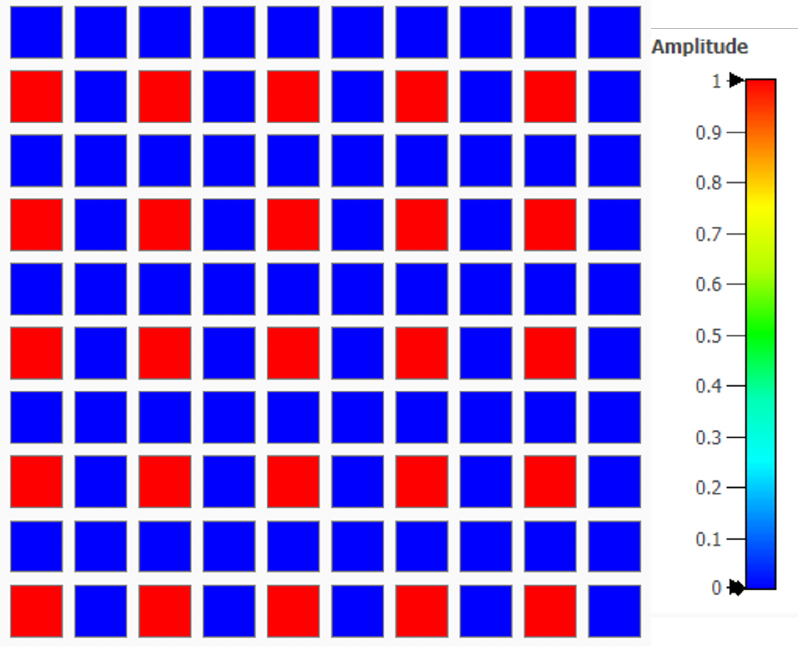}
\caption{RIS configuration to enable half-reflection.}
\label{fig:grating}
\end{minipage}
\hfill
\begin{minipage}[t]{0.325\linewidth}
\centering
\includegraphics[width=0.6\linewidth]{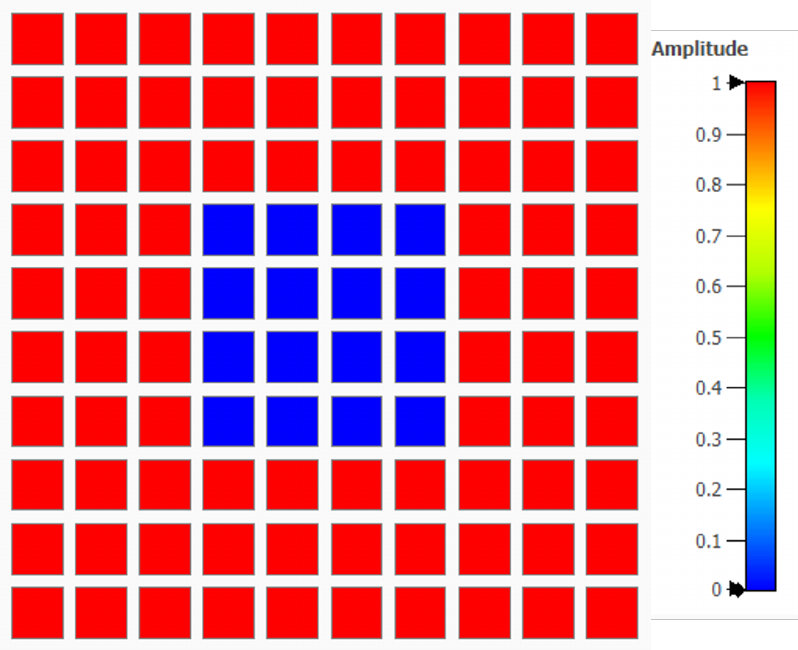}
\caption{RIS configuration to enable partial reflection.}
\label{fig:square}
\end{minipage}
%\vspace{-5mm}
\end{figure*}

\begin{figure*}[t]
\begin{minipage}[t]{0.325\linewidth}
\centering
\includegraphics[width=0.7\linewidth]{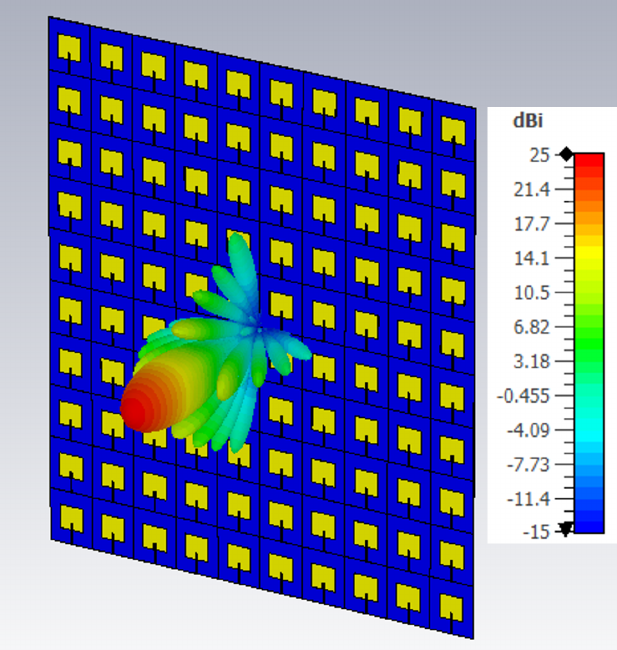}
\caption{Radiation diagram for full-reflective RIS.}
\label{fig:fullRefl_beam}
\end{minipage}
\hfill
\begin{minipage}[t]{0.325\linewidth}
\centering
\includegraphics[width=0.7\linewidth]{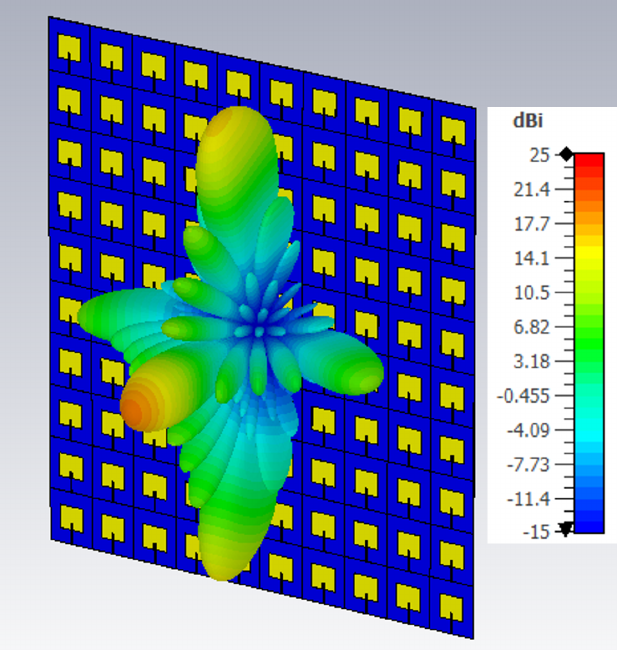}
\caption{Radiation diagram for half-reflective RIS.}
\label{fig:grating_beam}
\end{minipage}
\hfill
\begin{minipage}[t]{0.325\linewidth}
\centering
\includegraphics[width=0.7\linewidth]{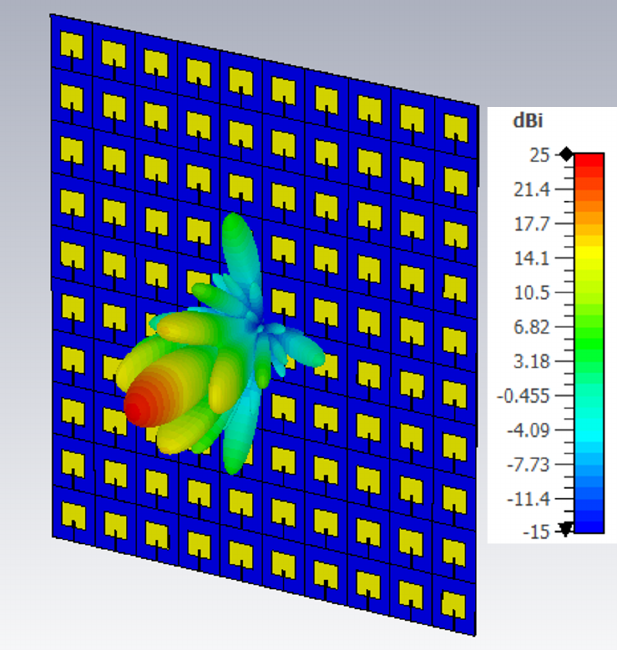}
\caption{Radiation diagram for partially-reflective RIS.}
\label{fig:square_beam}
\end{minipage}
\vspace{-5mm}
\end{figure*}

\subsection{CST Simulations}

We analyze how to properly select a RIS configuration. We use a full-wave ray tracing software, namely CST \cite{cststudio}, to emulate a conventional planar RIS with $10\times 10$ antenna elements, operating at $5.3$GHz. In our experimental campaign, we assume a $1$-bit configuration (this would turn the optimization variable $\v$ in \emph{Problem 1~(SHIELD)}) into a discretized variable that can be $0$ or $1$): each RIS antenna element can reflect or absorb the impinging signal.

We evaluate three different configurations: $i$) all antenna elements are set to fully reflect the signal, as shown in Fig.~\ref{fig:fullRefl}, $ii$) half of the available antenna elements are set to fully reflect the signal, as shown in Fig.~\ref{fig:grating} and $iii$) a small square of antenna elements ($4\times 4$) in the middle of the RIS is set to absorb the signal, as shown in Fig.~\ref{fig:square}. For each configuration, we run our simulations and depict the corresponding radiation diagram.

It shall be noted that only in the second configuration case, the results show a grating lobes effect (as depicted in Fig.~\ref{fig:grating_beam}. While this may appear a drawback of our solution, it suggests interesting hidden aspects: random configuration may result in unwanted RIS reflections and behaviors. However, if grating lobes are taken into account, the RIShield framework can still optimally operate and reduce the EM field exposure for specific users and areas. Additionally, the first and the third case (as depicted in Figs.~\ref{fig:fullRefl_beam}-~\ref{fig:square_beam}), show a single main beam (with different amplitude values) suggesting that they can be deployed to accurately reduce the signal leakage.

\section{Conclusions}

We have presented RIShield, a novel and efficient solution for enabling electromagnetic blackout in radiation-sensitive environments. By leveraging RISs strategically placed in such environments, we have demonstrated the capability to control and manipulate electromagnetic waves effectively. 

The primary goal of RIShield was to create a reliable mechanism to mitigate electromagnetic interference and ensure the safety and proper functioning of radiation-sensitive equipment and systems. 
The success of our solution represents a significant step forward in electromagnetic interference management and radiation-sensitive equipment protection.
Our experimental results---using two different commercial tools---have shown that RIShield can significantly reduce the impact of external electromagnetic signals on sensitive areas creating a protective shield.

%\section*{acknowledgments}
%The research leading to these results has received funding from the EU FP for Research and Innovation Horizon 2020 under Grant Agreements No. 101017011 (RISE-6G), 861222 (MINTS) and 956256 (METAWIRELESS). 

%While our study has achieved promising results, we recognize that further research is needed to optimize and fine-tune RIShield's performance. Future investigations should focus on refining RIS configurations, optimizing transmit precoder, and exploring additional techniques to adapt the shielding process dynamically based on the environment's requirements.
%The success of our solution represents a significant step forward in electromagnetic interference management and radiation-sensitive equipment protection. %We envision RIShield playing a crucial role in enhancing the reliability, security, and efficiency of sensitive systems, ultimately benefiting society as a whole.

\bibliographystyle{IEEEtran}
\bibliography{references}

\end{document}